\documentstyle[sprocl,epsf]{article}

\def\Journal#1#2#3#4{{#1} {\bf #2}, #3 (#4)}

\def\NPA{{\em Nucl.\ Phys.\ }A}
\def\NPB{{\em Nucl.\ Phys.\ }B}
\def\PLB{{\em Phys.\ Lett.\ }B}
\def\PRL{\em Phys.\ Rev.\ Lett.}
\def\PR{{\em Phys.\ Rep.}}
\def\PRA{{\em Phys.\ Rev.\ }A}
\def\PRC{{\em Phys.\ Rev.\ }C}

\def\ANP{{\em Adv.\ Nucl.\ Phys.}}

\def\be{\begin{equation}}
\def\ee{\end{equation}}
\def\bea{\begin{eqnarray}}
\def\eea{\end{eqnarray}}

\newcommand{\ben}{\begin{eqnarray*}}
\newcommand{\een}{\end{eqnarray*}}
\newcommand{\sing}{{}^1\!S_0}

\newcommand{\Lnn}{\Lambda_{{\rm NN}}}
\newcommand{\as}{a}
\newcommand{\wt}{\widetilde}

\begin{document}

\title{CONVERGENCE ISSUES IN NUCLEON-NUCLEON EFFECTIVE FIELD THEORY}

\author{JAMES~V.~STEELE}

\address{Department of Physics, The Ohio State University\\ 
	 Columbus, OH 43210--1106, USA\\
	 E-mail: jsteele@mps.ohio-state.edu}


\maketitle\abstracts{
The origin of the breakdown scale in an effective field theory
treatment of nuclear forces is investigated.  
Different organizational schemes used in the nonperturbative
calculation of nucleon-nucleon scattering seem to break down for
different physical reasons.
It is argued that the same physics is actually responsible, namely
strong two-pion exchange as also observed in the pion scalar form
factor. 
The prospect of extending the applicability of the effective field
theory beyond this breakdown point is discussed.
}

\section{Introduction}

Strong interactions at low energies can be described by an
effective field theory (EFT) of pions and nucleons with the
nucleons treated nonrelativistically.~\cite{all} 
Chiral symmetry prescribes how pions interact.
All other interactions are parameterized by the most general local 
Lagrangian consistent with the symmetries of quantum chromodynamics (QCD).

The coefficients of these general terms can either be predicted
directly from QCD, 
which may be possible in future lattice calculations, or fit from
existing data.
I will focus on nucleon-nucleon (NN) scattering in the following.
The data here is quite good and should lead to an excellent
determination of these constants.  
However, since nuclear forces produce bound states, a 
nonperturbative treatment of the EFT is required, 
which adds a new twist to this procedure.~\cite{all}

Any EFT will break down at a scale associated with the underlying
physics not taken into account in the original Lagrangian.  For
nuclear EFTs, an optimist may believe this will occur when the center
of mass momentum of the nucleons excite either the exchange of the next
heaviest meson after the pion, $p\sim m_\rho$, or the nucleon-$\Delta$
mass difference $p=\sqrt{M (M_\Delta-M)}\sim 525$~MeV.
In practice, the breakdown scale seems to be much lower, $p\sim 300$~MeV.
This could be due to anything from quark substructure to 
subtleties in the fitting.

Furthermore, two different organizational schemes for calculations in
the EFT seem to imply different physical reasons for the breakdown.
This would be unfortunate, since alternative ways to organize the
corrections should have no effect on the physics.
By using some models, I will show what actually does set this scale
and then use this to make connections to the real world.

Extending the applicability of the EFT to the scale associated with
the $\Delta$ seems like a small achievement, but could make
calculations of nuclear matter converge much faster. 
In that case, the momentum of interest 
is on the order of the Fermi momentum at saturation
$p_F\sim 280$~MeV, which
currently is too close to the breakdown scale for an EFT analysis
to be useful.

\section{Removing vs. Weakening the Pion}\label{rm}

Assume, for now, that interactions between two nucleons are exactly
given by potential exchange of both a long- and short-ranged meson
with masses $m_\pi=140$~MeV and $m_\rho=770$~MeV respectively,
\be
V_{\rm exact}(r) = - \alpha_\pi \frac{e^{-m_\pi r}}{r} - \alpha_\rho
\frac{e^{-m_\rho r}}{r}\, .
\label{yuks}
\ee
The pion coupling is taken to be
$\alpha_\pi=g_A^2 m_\pi^2/(16 \pi f_\pi^2)\simeq0.075$,
as in the real world, and the rho coupling
$\alpha_\rho=1.05$ is tuned to give a large scattering length
$\as=-23.4$~fm, as observed in data for 
the $\sing$ partial wave of NN scattering.  
The following discussion does not depend on the value of the potential
at the origin, so I will take $V_{\rm exact}$ to be zero there for
simplicity.

\begin{figure}
\begin{center}
\leavevmode
\epsfxsize=2.3in
\epsffile[84 621 201 725]{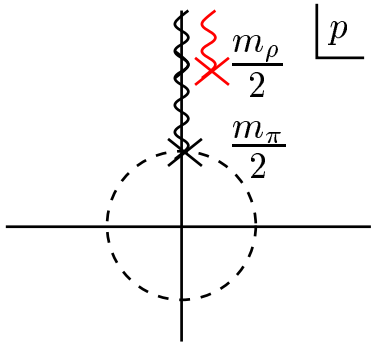}
\end{center}
\caption{\label{cut}The analytic structure of an observable for the
two-Yukawa model of Eq.~(\protect\ref{yuks}) in the complex momentum
plane.} 
\end{figure}

This potential generates observables with analytic structure as shown
in Fig.~\ref{cut}, specified by a cut at both $m_\pi/2$ and $m_\rho/2$.
The existence of these cuts indicates that
a low-momentum expansion of an observable will not converge
for momenta above $m_\pi/2$.  
An EFT can only go beyond this point if it either exactly reproduces
or sufficiently approximates the analytic structure of the
pion.

This can be illustrated by focusing on the most general Lagrangian without
pions (I will restrict myself to the $\sing$ channel below) 
\be
{\cal L}_{\rm EFT} = N^\dagger \left[ i \partial_t + \frac{\nabla^2}{2M}
\right] N
- C_0 (N^\dagger N)^2 + \frac12 C_2 \left[ (N^\dagger
\nabla N)^2 + h.c. \right] + \ldots \, ,
\label{Leff}
\ee
which corresponds to a nonrelativistic potential between two nucleons of
\be
V_{\rm EFT}(p,p') = C_0 + \frac12 C_2 \left( p^2 + p'{}^2\right) +
\ldots \, ,
\label{Veff}
\ee
where $p$ and $p'$ are the center of mass momentum of the incoming and
outgoing nucleons.
Since these interactions depend on powers of momentum,
a naive calculation of the NN scattering phase shifts from
${\cal L}_{\rm eff}$ leads to ultraviolet divergences.
This just reflects our ignorance of the high-energy
behavior of this theory.

Two possible procedures to regulate these divergences 
are: i) adding a cutoff to all integrals to eliminate or suppress the
high momentum states and ii) using dimensional regularization, which
amounts to throwing out any integrals that diverge like powers of
momentum. 
The second procedure may seem a bit counterintuitive at
first, 
but is simpler to implement since it does not break any
symmetries of the underlying theory and therefore does not require
additional counterterms.
Perturbative calculations in chiral perturbation theory
have confirmed the equivalence of these regularization procedures.

Nonperturbative calculations, on the other hand, show that a naive
implementation of dimensional regularization is incomplete.
A generalization to account for integrals that linearly diverge
is required.
This regularization is called power divergence subtraction~\cite{PDS}
(PDS).

After regulating the theory, the EFT can be organized to give 
predictions order-by-order in powers of an expansion
parameter. 
Since the low-energy constants $C_{2n}$ are not determined by the symmetries
of QCD, they must be fit to data.
In the process, they will typically pick up the next scale $\Lambda$ of 
physics not explicitly accounted for in the EFT, leading to
$C_{2n} p^{2n} \sim (4\pi/\Lambda^2) (p/\Lambda)^{2n}$.
Therefore interactions with more derivatives are suppressed by additional
powers of $p/\Lambda$.
Power counting in this parameter
allows Eq.~(\ref{Leff}) to be truncated to the first few terms, with
more accurate results obtained by keeping more terms in the
momentum expansion.

There are two prevalent ways of implementing power counting
in nonperturbative effective field theories.  
The first is to count powers of $\Lambda$ in the potential 
$V_{\rm EFT}$ 
and then solve the Lippmann-Schwinger equation.
Such a procedure sums an infinite number of Feynman diagrams to
generate the nonperturbative amplitude ${\cal A}$, related to 
the phase shift $\delta$ by
\be
{\cal A} = \frac{4\pi/M}{p\cot\delta-ip} \, .
\ee
This method of power counting is called
$\Lambda$-counting~\cite{lepage} and is in the spirit of 
the Hamiltonian theory of Wilson, first applied to nuclear EFTs by
Weinberg.~\cite{Weinberg} 
An alternative is to expand ${\cal A}$ in powers of $Q\equiv
(p,m_\pi,1/\as)$, motivated by chiral perturbation theory, which
expands in $p$ and $m_\pi$.
The nonperturbative nature of the interactions identifies $1/\as$ as a
small parameter as well, justifying its inclusion in the expansion.
This method of power counting is called $Q$-counting.~\cite{PDS,cohen}

For a theory without pions, such as Eq.~(\ref{Leff}), the underlying
scale $\Lambda$ is just $m_\pi$.
Both power counting procedures
work as long as momenta are less than this scale,
reproducing the effective range expansion 
\be
p\cot\delta = - \frac1{\as} + \frac12 p^2 \sum_{n=0}^\infty r_n
\frac{p^{2n}}{\Lambda^{2n}} \, . 
\label{ERe}
\ee
The effective ranges, which are determined from NN scattering
data,~\cite{NNdata} also exhibit this scale
$r_n\sim 1/\Lambda\sim 1/m_\pi$.
To extend the EFT to higher momenta, 
pions need to be explicitly added to the effective
Lagrangian.
This hopefully will bring $\Lambda$ up to the QCD scale of
$m_\rho$ or possibly even 1~GeV.

The inclusion of pions nonperturbatively 
is different in the two power counting procedures.
Every iteration of one-pion exchange contributes to the pion cut in
Fig.~\ref{cut}. 
In general, $n$ iterations are proportional to%
\footnote{This can be seen by studying the exact solution of an
exponential potential,~\cite{us2}
which has the same cut structure as a Yukawa
or by looking at the analytic expressions of higher Yukawa loops,
which exhibit the 
cut structure through Polylogarithms.~\cite{binger}}
\be
\mbox{$n$-Pion Exchange} \propto
\left[ \frac{m_\pi}{\Lnn}
\frac{m_\pi^2}{4 p^2} \ln \! \left( 1 + \frac{4 p^2}{m_\pi^2} \right)
\right]^n\, ,
\label{ope}
\ee
where $\Lnn\simeq300$~MeV is just another way of expressing the strength 
of the interaction via the relation $\alpha_\pi=m_\pi^2/(M\Lnn)$.
$\Lambda$-counting includes this long-range part of one-pion exchange
fully at leading order, removing the effect of the pion cut.
$Q$-counting pushes one-pion exchange
to subleading order and iterated one-pion exchange to even higher
orders. 
Therefore, each order in the $Q$-counting expansion 
weakens the pion cut by an additional power of $m_\pi/\Lnn$, 
as seen in Eq.~(\ref{ope}).
The topic of the next section is whether 
this simpler implementation of power counting, also called
perturbative pions, can be as
successful as $\Lambda$-counting.

\section{New Fitting Procedures}\label{fits}

The NN EFT is expected to work best at low-momentum.
So a natural
procedure to fix the low-energy constants $C_{2n}$ 
is to calculate an observable, like $p\cot\delta$, and fit
near threshold.
However, there are some complications originating
from the nonperturbative nature of the problem.  
These must be cleared up before being able to make a statement about
the breakdown scale.

\begin{figure}
\begin{center}
\leavevmode
\hbox{
\hspace{-1.25cm}
\epsfxsize=2.5in
\epsffile{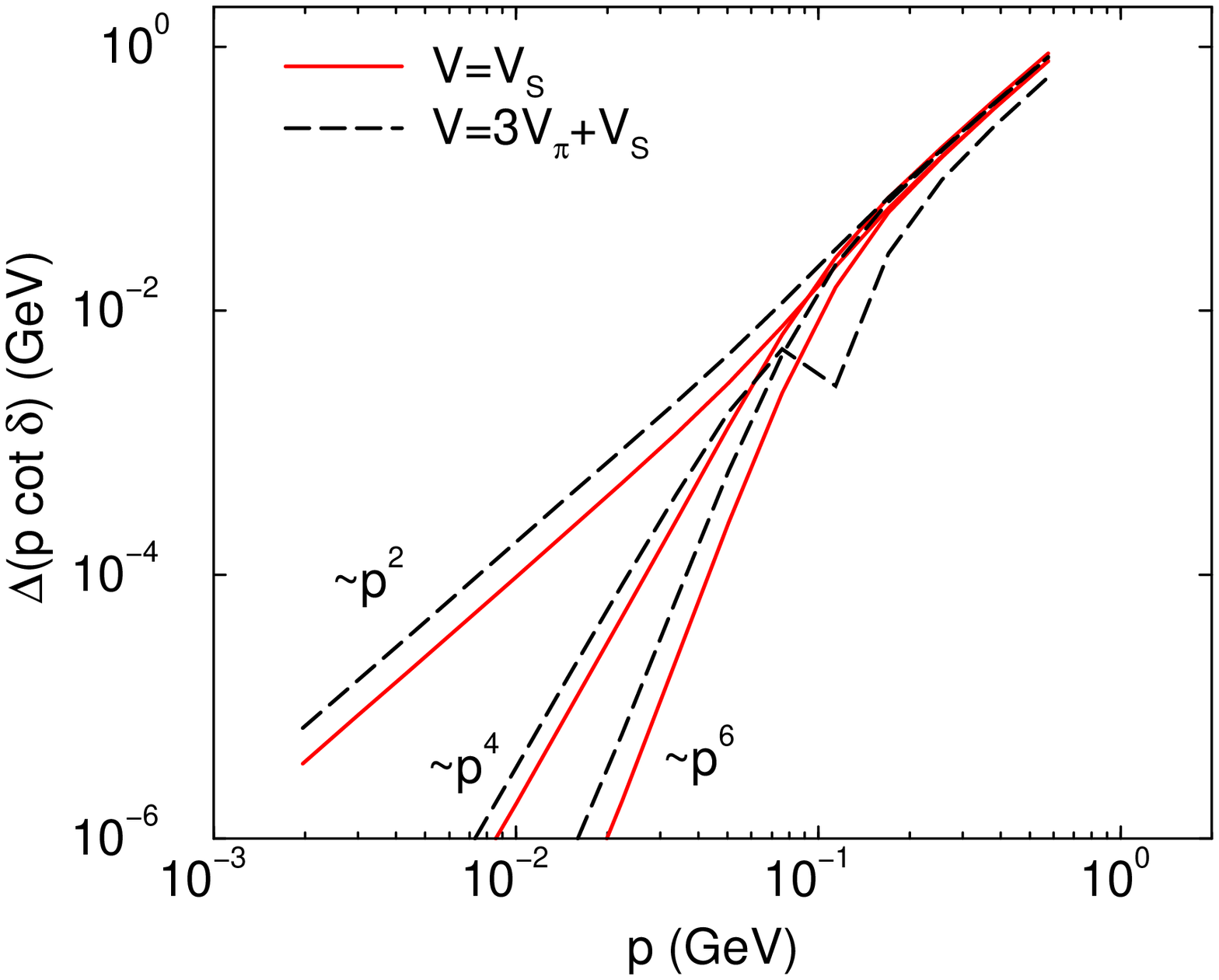}
\hspace{.25cm}
\epsfxsize=2.5in
\epsffile{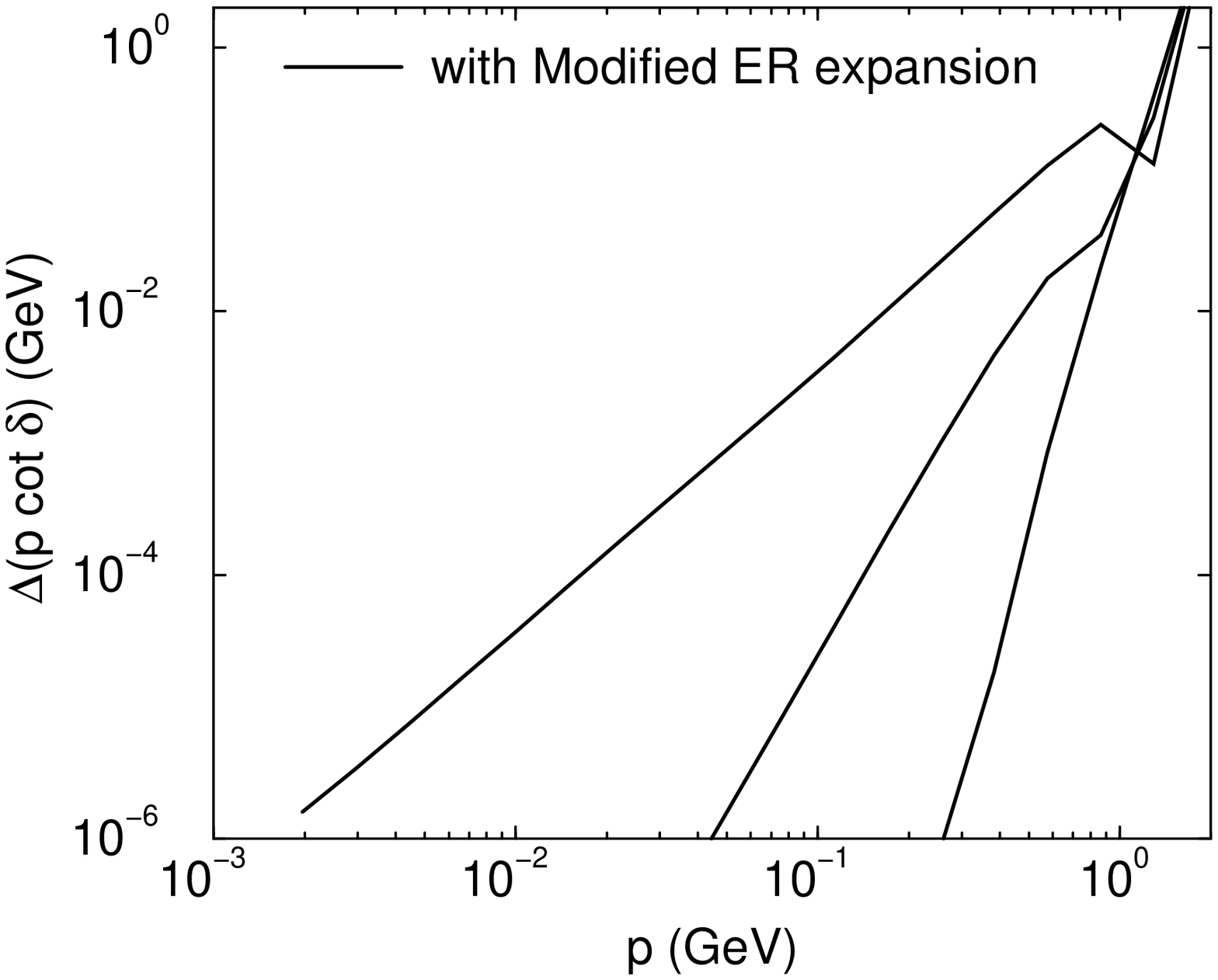}
}
\end{center}
\caption{\label{expmod}
The error in $p\cot\delta$ for a pion coupling three times stronger
than the real world. 
The left plot is from a normal fit to the low-energy constants both
without (solid) and with (dashed) explicit pions.
The right plot uses the modified effective range
expansion to fit.}
\end{figure}

\subsection{$\Lambda$-counting and the Modified Effective Range Expansion}

Most calculations that implement $\Lambda$-counting require a numerical
treatment (although there are exceptions~\cite{cohen}).  
A nice way to show the systematic improvement expected of an EFT in
this case is to use error plots~\cite{lepage,us1,us2} as in
Fig.~\ref{expmod}. 
As more and more constants are added to the effective potential
Eq.~(\ref{Veff}), the error in $p\cot\delta$
improves systematically by powers of
$p^2$, reflected in the improving slopes.   
The point at which all errors are of the same order is where the
theory is expected to break down.~\cite{us1}
The solid lines of the left plot of Fig.~\ref{expmod} show the case
without explicit pions.
The breakdown point, or radius of convergence, of the EFT is around
$m_\pi/2$, as expected from the discussion in Section~\ref{rm}.

Adding pions to the short-ranged Lagrangian Eq.~(\ref{Leff}) 
as dictated by chiral symmetry should improve the radius of
convergence.~\cite{bira} 
However, one needs to be careful when implementing this with cutoff
regularization. 
If the pion coupling were three times
stronger, for example, adding it to the Lagrangian would {\it not} improve
the breakdown scale.~\cite{us2} 
This is illustrated by the dashed lines in the left plot of
Fig.~\ref{expmod}.
Chiral symmetry helps in that the coupling is small, 
but an EFT should work
regardless of the strength of the potential being added.  
What went wrong?

\begin{figure}
\begin{center}
\leavevmode
\epsfxsize=5in
\hbox{
\hspace{-.5in}
\epsffile[121 578 538 674]{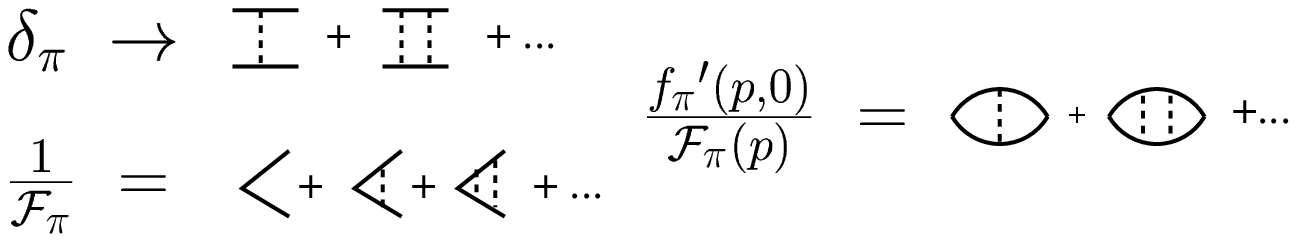}
}
\end{center}
\caption{\label{pions}Symbolic definition of quantities in the
modified effective range function in Eq.~(\protect\ref{MERF}), which
sums all the pion contributions.} 
\end{figure}

Note that the effective ranges $r_n$ of Eq.~(\ref{ERe}),
which are used to fix the values of the $C_{2n}$,
exhibit the scale $m_\pi$.
If this scale were completely removed from the fitting procedure, then the
constants should take on the proper scale $m_\rho$.
One way to do this is to construct a function similar to
$p\cot\delta$ which is analytic all the way to the rho cut in
Fig.~\ref{cut}. 
In proton-proton scattering analyses, such 
a procedure is used to remove the threshold cuts coming
from photon exchange.
Called the modified effective range expansion,~\cite{modER}
it can be used here to remove the effect of pion exchange by defining
\be
K(p) \equiv \frac{p\cot(\delta-\delta_\pi)}{|{\cal F}_\pi(p)|^2}
+ \mbox{Re}\, \left[ \frac{f_\pi'(p,0)}{{\cal F}_\pi(p)} \right] \equiv
-\frac1{\wt\as} + \frac12 p^2 \sum_{n=0}^\infty \wt{r}_n
\frac{p^{2n}}{\wt{\Lambda}^{2n}} \ .
\label{MERF}
\ee
All possible insertions of one-pion exchange are taken into account
in this formula, as seen pictorially in Fig.~\ref{pions}.
For the model potential Eq.~(\ref{yuks}), this
function is analytic up to $m_\rho/2$ and has modified effective
ranges $\wt{r}_n\sim 1/\wt{\Lambda} \sim 1/m_\rho$.
Fitting the $C_{2n}$'s using the function
$K(p)$ leads to much
better results, shown in the right plot of Fig.~\ref{expmod}.
The radius of convergence is now on the order of $m_\rho$ as expected.

Using a cutoff to regulate the theory mixes long- and short-distance
physics, which is what led to the premature breakdown.~\cite{us2} 
The modified effective range expansion removes the
long-distance physics from the observable, allowing a clean fit of
the low-energy constants to the short-distance physics.
In principle, this can also be achieved by the addition of more counterterms,
but the modified effective range expansion is a simpler alternative.
Luckily, one-pion exchange in the real world is weak enough that an
improper treatment of this malady does not make a difference.~\cite{us2}
However, this might not be the case for two-pion exchange as will be
discussed in Section~\ref{2pi}.

\begin{figure}
\begin{center}
\leavevmode
\epsfxsize=4in
\epsfysize=3.7in
\epsffile{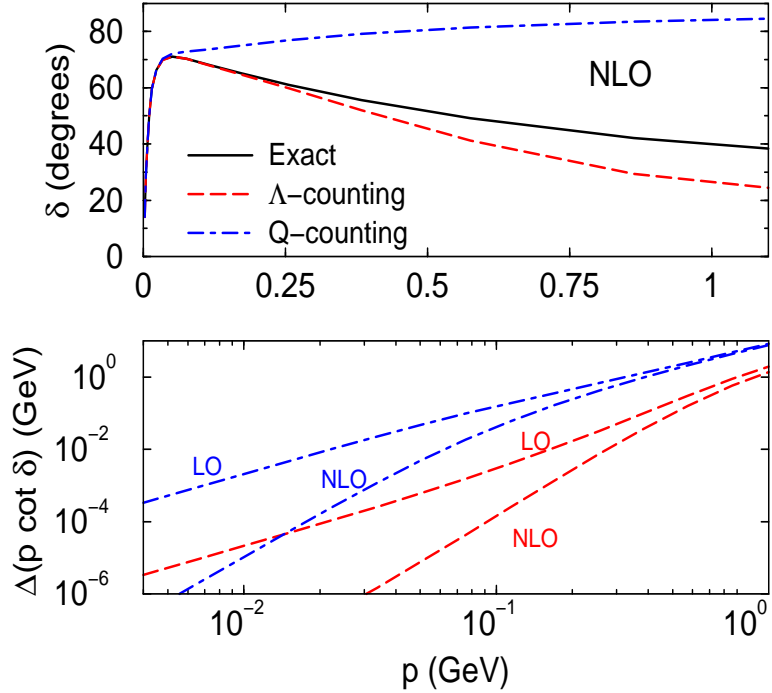}
\end{center}
\caption{\label{qcount}Comparison of the two power counting procedures
for the two-Yukawa model Eq.~(\protect\ref{yuks}) using a fit at low
momentum.} 
\end{figure}

\subsection{$Q$-counting and a Global Fit}

A low-momentum fit to data using
$Q$-counting and perturbative pions 
also breaks down much earlier than expected.~\cite{us2}
In Fig.~\ref{qcount}, the leading order (LO) and
next-to-leading order (NLO) calculations using $Q$-counting are compared to
those using $\Lambda$-counting.
The $\Lambda$-counting result has a radius of convergence around
$m_\rho$ after the low-energy constants are matched to the modified
effective range expansion.

Implementing the modified effective range expansion 
is inconsistent with the idea of perturbative pions, and so
the $Q$-counting result breaks down around $m_\pi$. 
However, it is important that the low-energy constants only 
pick up the (unknown) short-distance physics, and matching to 
the modified effective
range expansion is the most convenient way to do this.
Using PDS, this would lead to the expressions~\cite{kap}
\be
C_0 = \frac{4\pi/M}{-\mu+1/\wt{\as}} \, ,
\qquad
C_2 = \frac{MC_0^2}{4\pi} \frac12 \wt{r_0} \, ,
\qquad\ldots\, ,
\label{modPDS}
\ee
which only depend on the short-distance physics.%
\footnote{Although the renormalization scale $\mu$ appears in these low-energy
constants, all observables are independent of it.}
One important modification to this 
would be to match the pole of the amplitude ${\cal
A}$ exactly~\cite{mehstew} 
\be
C_0 = \frac{4\pi/M}{-\mu+1/\as} \, .
\label{c0}
\ee
Otherwise, terms attempting to correct the position of the pole 
would enter as double poles at higher orders messing up the power counting.
This modification requires perturbative corrections to $C_0$ that
compensate the pion contributions at each order.
However, this still leaves 
the other $C_{2n}$'s dependent on the modified effective
ranges $\wt{r}_n$, which cannot be obtained using perturbative pions.

\begin{figure}
\begin{center}
\leavevmode
\epsfxsize=3in
\epsffile{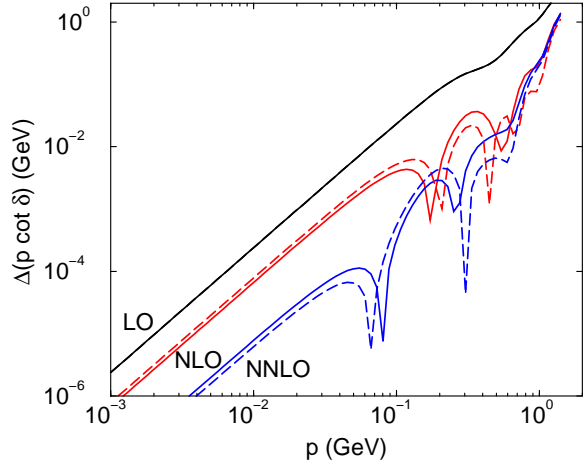}
\end{center}
\caption{\label{toy}$Q$-counting expansion of the two delta-shell
model.  The exact expression (solid) coincides nicely with 
the EFT (dashed) when a global fit is minimized with respect to 
the modified effective ranges.  The dashed line at LO is on top of the
solid line.}
\end{figure}

A method must therefore be devised to carry out the matching of
Eq.~(\ref{modPDS}).
It is reasonable to assume that when the modified effective ranges
are set to the correct values, the error of the EFT will be less than when
the ranges are at some random values.
This can be checked by using a model.
Replacing the Yukawas in Eq.~(\ref{yuks}) by delta-shells
\be
- \alpha_i \; \frac{e^{-m_i r}}{r} \to
- g_i \; \delta\!\left(r-\frac1{m_i} \right) \, ,
\ee
an analytic expression for the exact answer can be obtained.~\cite{kap}
Expanding the exact amplitude in powers of $Q$, it can be compared
to the blind EFT fit of the low-energy constants to data 
using $Q$-counting.
The results coincide best when the modified effective ranges
are varied to globally minimize the $\chi^2$ for data in a
representative momentum range like $[m_\pi,m_\rho]$.

The results are shown up to next-to-next-to-leading order (NNLO) in
Fig.~\ref{toy}. 
They suggest that for $Q$-counting, a constrained global fit of the EFT
to data to fix the low-energy constants is better than 
a low-momentum fit.
The constraint is that the modified effective
ranges, not the low-energy constants themselves, are varied to achieve
the best fit.

One thing to notice about Fig.~\ref{toy} is that the errors do not
show the systematic improvement in slope characteristic of the earlier
error plots.
Although the scattering length is fit exactly to
reproduce the pole in the amplitude Eq.~(\ref{c0}), 
the other terms in the effective
range expansion have corrections at each order in the $Q$-counting.
A comparison of Fig.~\ref{toy} with the right plot of
Fig.~\ref{expmod} shows that 
accuracy is sacrificed near threshold to obtain the global fit needed
for this power counting.
In addition, the ability to read the EFT radius of convergence 
from the error plot is gone.

\begin{figure}
\begin{center}
\leavevmode
\epsfxsize=3.5in
\epsffile{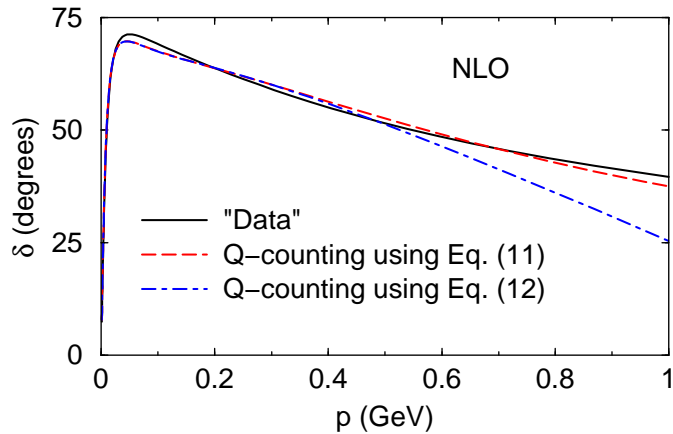}
\end{center}
\caption{\label{2yuk}Determination of the $Q$-counting breakdown scale
for the two-Yukawa model.}
\end{figure}

Another way to estimate the breakdown scale is therefore needed.
The calculated amplitude ${\cal A}$ starts at order $Q^{-1}$ and has
a contribution ${\cal A}_n$ at each order $Q^n$ in the expansion. 
Up to this point, I have extracted the $Q$-counting phase shift 
by expanding the expression~\cite{PDS}
\bea
\delta &=& 
\frac1{2i}\; \ln\!\left[ 1 + i \frac{Mp}{2\pi} \left( {\cal A}_{-1} +
{\cal A}_0 + \ldots \right) \right] 
\label{exp1}
\\
&\to& \frac1{2i}\; \ln\!\left( 1 + i \frac{Mp}{2\pi} {\cal A}_{-1}
\right) + \frac{Mp}{4\pi} \left( \frac{{\cal A}_0}{1+i \frac{Mp}{2\pi}
{\cal A}_{-1}} \right) + \ldots
\, ,
\label{exp2}
\eea
but I could also have kept the full expression Eq.~(\ref{exp1}).
These two expressions are equivalent up to terms that are
higher order in the $Q$ expansion.
Those extra terms, though, are an estimate of the corrections to the
actual result and depend on the radius of convergence.
The breakdown for $Q$-counting can be identified as the
point at which the two expressions Eqs.~(\ref{exp1}) 
and (\ref{exp2}) diverge from each other. 
Doing this exercise for the two-Yukawa model results in
Fig.~\ref{2yuk}.
There it is easy to see the actual breakdown is somewhere around
$m_\rho$ as expected.

\section{Data and the Role of Two-Pion Exchange}\label{2pi}

\begin{figure}
\begin{center}
\vspace*{-.3cm}
\leavevmode
\hbox{
\hspace{-1.25cm}
\epsfxsize=2.5in
\epsffile{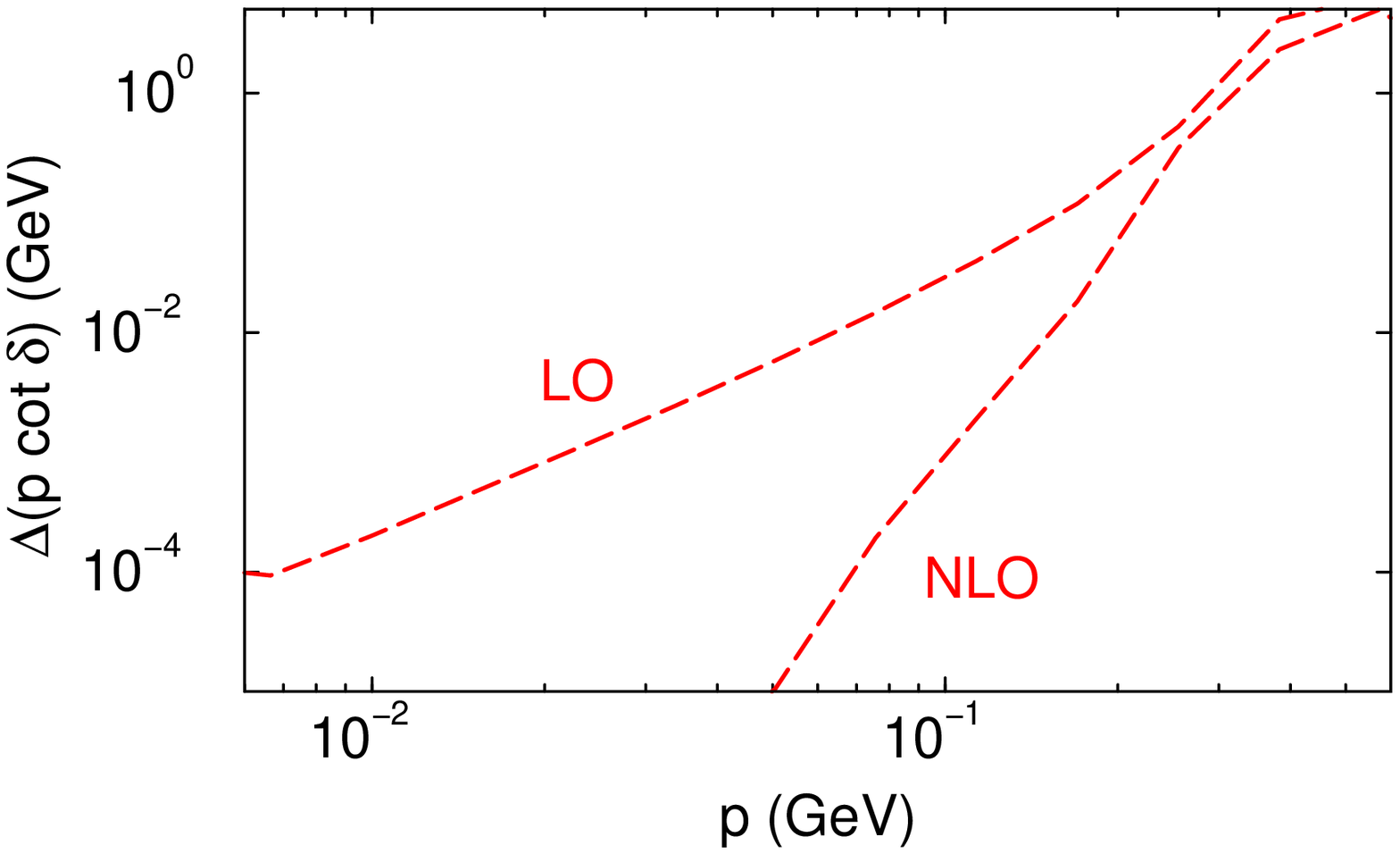}
\hspace{.25cm}
\epsfxsize=2.5in
\epsffile{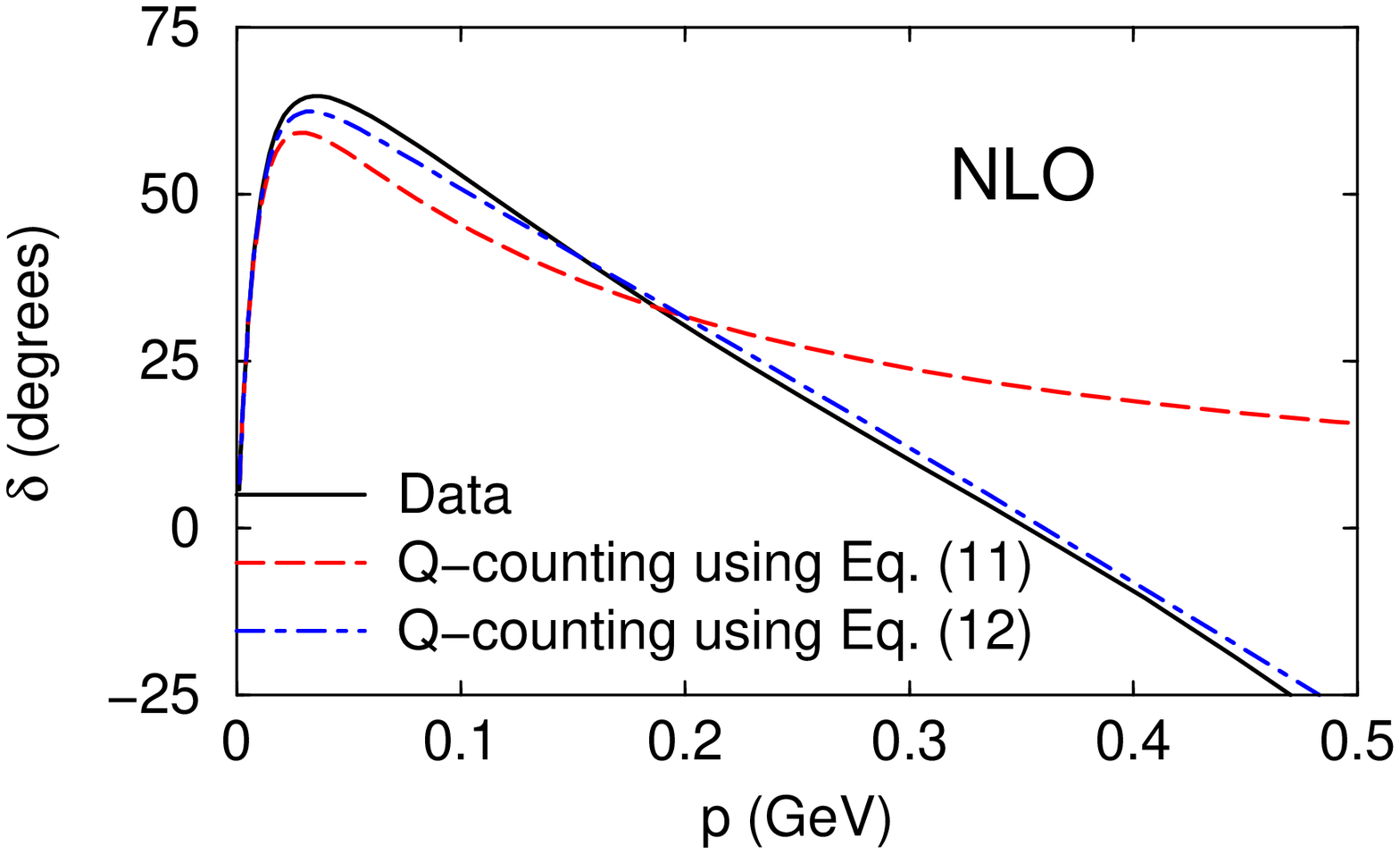}
}
\end{center}
\caption{\label{data}
Results for $\sing$ NN-scattering data.
The breakdown scale is determined by the left plot for
$\Lambda$-counting and the right plot for $Q$-counting.}
\end{figure}

Now that the two power counting methods are well understood and give
the proper radius of convergence for a model, it is appropriate to
apply them to actual NN-scattering data.
Using the Nijmegen partial wave analysis,~\cite{NNdata}
the radius of convergence for each of the two different power
counting schemes is shown in Fig.~\ref{data}.
The left plot shows that, for $\Lambda$-counting with one-pion
exchange included as in the previous section, the breakdown is around
$p\sim 300$~MeV.
The right plot shows a similar scale for
$Q$-counting with perturbative pions.

This scale is less than expected and seems to be set by physics other
than the $\rho$ or $\Delta$.
Early calculations in $Q$-counting predicted a breakdown
at $\Lnn$, which would coincide with the observed scale.
However, recall that in the previous section I showed that
$Q$-counting broke down around $m_\rho$ in the two-Yukawa model.
That calculation had identical content with respect to one-pion
exchange, indicating that the $300$~MeV comes from different physics.
A closer study of $\Lambda$-counting determines that two-pion
exchange should be included at NLO.~\cite{bira}

The fact that the actual data behaves so much differently from the
two-Yukawa model indicates the model was not sophisticated enough.
Furthermore, no reasonable choices for the parameters
of this model can give as large an effective range
$r_0=2.63$~fm as seen in the $\sing$ channel.~\cite{us2} 
Until now I have assumed that $m_\pi$ and $m_\rho$ are the underlying
scales, but nuclear physics wisdom tells us there are actually two
short-range Yukawas
\be
V(r) = - \alpha_\pi \frac{e^{-m_\pi r}}{r} 
       - \alpha_\sigma \frac{e^{-m_\sigma r}}{r} 
       + \alpha_\rho \frac{e^{-m_\rho r}}{r} \, ,
\label{3yuks}
\ee
given by an attractive scalar (of mass between $500$ and $600$~MeV)
and repulsive vector particle.%
\footnote{The vector particle is normally taken to be the $\omega$
meson, but for continuity with the earlier discussion, I will call it
the $\rho$ meson.}
Taking the short-distance couplings to be much stronger than
the pion $(\alpha_\sigma,\alpha_\rho)=(7,14.65)$,
the modified effective ranges $\wt{r}_n$ of the data can also
be reproduced.~\cite{us2}
This three-Yukawa model is reminiscent of the Bonn
potential,~\cite{mach} which is known to model the data well.

Applying the NN EFT to observables produced from Eq.~(\ref{3yuks})
indeed leads to a breakdown around $m_\sigma/2\sim300$~MeV.
I can verify the breakdown is not set by the two-pion threshold
$2m_\pi$ within this model by varying the pion and scalar masses.
The $\sigma(600)$ particle is a common parameterization in nuclear
physics for two-pion exchange contributions.  
The large scalar coupling indicates that including
such effects through chiral symmetry
might require the modified effective range expansion.
Such an analysis could improve the EFT radius of convergence.%
\footnote{A preliminary calculation, inspired by discussions at 
this workshop, seems to indicate this is true.~\cite{us3}}

It might seem odd that two-pion effects could be strong, since
they should be weak according to chiral symmetry.~\cite{lepage}
However, there are other instances where two-pion effects have
proven to be strong.
The pion scalar form factor, defined by
\be
\langle \pi^a(p) \pi^b(p') | {\hat m} \left( \bar{u} u + \bar{d} d \right)
| 0 \rangle = \delta^{ab} \sigma_\pi \Gamma(s), \qquad\qquad
s= (p+p')^2,
\ee
has been shown to have poor convergence in chiral perturbation
theory,~\cite{meiss} breaking down just above threshold at
$\sqrt{s}\simeq 400$~MeV.
This comes about from infrared effects that require
the summation of $(\ln s)^n$ terms to all orders.
Carrying out this summation does
lead to improved results and the generation of 
a ``resonance''-like shape in the dispersion analysis just
like a scalar resonance around $600$~MeV.~\cite{meiss}
The same phenomenon could occur in NN scattering and requires
further study.

\begin{figure}
\begin{center}
\leavevmode
\epsfxsize=3in
\epsffile{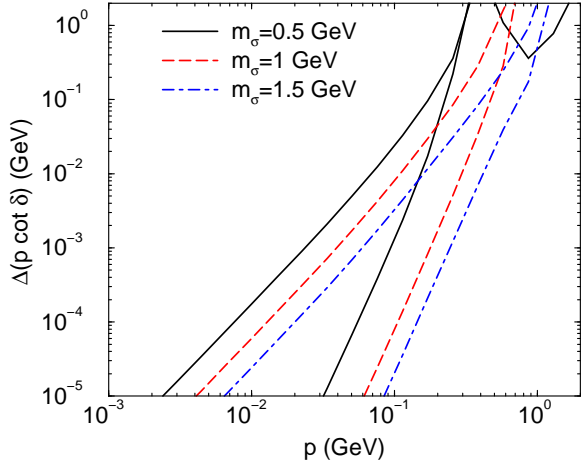}
\end{center}
\caption{\label{sig}The three-Yukawa model Eq.~(\protect\ref{3yuks})
with different values for the scalar mass, but $\as$, $r_0$, and
$m_\rho-m_\sigma$ fixed.  This shows that the radius of convergence
tracks with the changing $m_\sigma$.} 
\end{figure}

\section{Red Herrings}

The models developed to understand the breakdown scale
better can also be used to shed light on other convergence issues.
One feature of the three-Yukawa model is that it can reproduce the
somewhat large effective range in NN scattering, which is
$r_0\sim2/m_\pi$.
Any reasonable choices of parameters for the two-Yukawa model can only
give $r_0$ around $1/m_\pi$.
Therefore, it is possible that
the lower radius of convergence of the three-Yukawa model
might have nothing to do with two-pion effects, but come from
some type of correlation with the effective range $r_0$ instead.
Choosing different parameters in Eq.~(\ref{3yuks}), 
I can keep $\as$, $r_0$, and the mass difference $m_\rho-m_\sigma$ fixed 
while changing $m_\sigma$ from $500$~MeV to $1500$~MeV.
The results are shown in Fig.~\ref{sig}, confirming that the radius of
convergence is set by $m_\sigma/2$ alone.

Recently, the convergence of $Q$-counting has been challenged within
the context of a low-energy theorem.~\cite{cohen}
The point can be made by defining the observable
\be
S(p^2) \equiv p\cot\delta - \left( - \frac1{\as} + \frac12 r_0 p^2 
\right)\, ,
\ee
and taking $p=|1/\as|$.
Then $Q$-counting has two types of corrections:
\be
\frac1{\as \Lnn} \ll 1\, , 
\qquad\qquad\qquad 
\frac{m_\pi}{\Lnn} \sim \frac12\, .
\label{exppar}
\ee
It was shown that the $Q$-counting prediction for $S(p^2)$
at NLO is very far from the data.

This can also be seen in a model.  
For an analytic understanding, I will use the two delta-shell
potential of Section~\ref{fits}.
Expanding in $Q$, 
\bea
S\left(\frac1{\as^2}\right) &=& \left( -22.9+40.2-23.3+7.9+
\ldots\right) \times 10^{-7}  m_\pi\, ,\\
&&\quad \mbox{\sc nlo} \quad\; \mbox{\sc nnlo} \quad \mbox{\ldots}
\nonumber
\eea
one finds the NNLO term is larger than the NLO term.
Only after a few more orders does the series start to converge.

Taking the exact answer, I can also expand in $Q$ to see what is
happening analytically.
These expressions are messy and can be simplified by observing
that even though $m_\pi\as ={\cal O}(1)$ in $Q$-counting, it is actually
of order 10 numerically.
Expanding $S(1/\as^2)$ in $(m_\pi \as)^{-1}$ then should 
simplify the expression immensely without blurring its
meaning, leading to
\bea
S\left(\frac1{\as^2}\right) &=& \frac{m_\pi}{(m_\pi \as)^4} \left[ 
- \frac{m_\pi}{3 \Lnn} 
+ \frac{m_\pi^2}{\Lnn^2} \left( \frac{13}{15} +
\frac{8 \Lnn}{9 m_\rho} \right)
+ \ldots \right]
\\
&&{}+ \frac{m_\pi}{(m_\pi \as)^5} \left[ 
\frac{4 m_\pi}{15\Lnn} 
-\frac{m_\pi^2}{\Lnn^2} \left( \frac{68}{45} + \frac{4\Lnn}{9m_\rho}
\right)
+ \ldots \right] + \ldots\, .
\eea
Terms in the second bracket and higher are
suppressed by $(m_\pi \as)^{-1}$ and can be neglected.
Focusing on the first bracket of terms, it is clear that the expansion
is in $m_\pi/\Lnn$ as expected for $Q$-counting.
However, the leading term (which happens to be NLO for this
particular observable) has a numerical factor of $\frac13$, which
suppresses it down to the order of the NNLO term.
Therefore, the $Q$ expansion of $S(p^2)$
works, but is slowly converging due to
some inconvenient numerical factors as well as the size of the
expansion parameter, Eq.~(\ref{exppar}).

\section{Summary}

Nonperturbative effective field theories are subtle to implement.
Techniques used successfully in chiral perturbation theory, such as
naive dimensional regularization and low-momentum fitting, do not
produce the expected results in nonperturbative calculations.
After a careful study using solvable models, new methods have been
developed to deal with these complications.~\cite{PDS,us2,kap}

Pions enter the calculation differently depending on whether power
counting is done in the QCD scale ($\Lambda$-counting) or the chiral
scale ($Q$-counting). 
In both cases a proper determination of the low-energy constants from the
short-distance physics requires more than just a low-momentum fit.
Utilizing the modified effective range expansion to remove the effect
of the long-distance pions from the process of
fitting, the $\Lambda$-counting
radius of convergence moves beyond the pion scale.~\cite{us2}
A constrained
global fit minimized with respect to the modified effective ranges
is required to achieve the same effect in $Q$-counting.~\cite{kap}

Applying the new fitting procedures to $\sing$ NN scattering
data results in a breakdown of about $300$~MeV.
Models can be used to confirm this breakdown is associated with the
same physics for both power counting procedures.
Using a three-Yukawa model, it was argued that including two-pion
exchange will extend the EFT radius of convergence.
It is not unreasonable to assume this effect is strong since a similar
situation occurs in the pion scalar form factor.~\cite{meiss}

Finally, the convergence of $Q$-counting was investigated using an
exactly solvable model.  
It was shown that some observables could have inconvenient numerical
factors, such as occur in the low-energy theorem,~\cite{cohen}  which
suppress the leading orders and mask the behavior of the expansion.
However, analytical results show that this expansion does
converge, albeit slowly.

\section*{Acknowledgments}
The work presented in this talk was done in collaboration with
R.~J.~Furnstahl and D.~B.~Kaplan.  
I would like to thank the organizers, P.~Bedaque,
M.~J.~Savage, R.~Seki, and U.~van Kolck, for an
enjoyable and productive workshop
and the Institute of Nuclear Theory at the University of Washington
for its hospitality.
This work was supported by the National Science Foundation
under Grants No.\ PHY--9511923 and PHY--9258270.

\end{document}